\input harvmac
%\draftmode
\def\journal#1&#2(#3){\unskip, \sl #1\ \bf #2 \rm(19#3) }
\def\andjournal#1&#2(#3){\sl #1~\bf #2 \rm (19#3) }

\def\ie{{\it i.e.}}
\def\eg{{\it e.g.}}

\def\frac#1#2{{#1\over#2}}

\def\inbar{\,\vrule height1.5ex width.4pt depth0pt}
\def\IC{\relax\hbox{$\inbar\kern-.3em{\rm C}$}}
\def\IR{\relax{\rm I\kern-.18em R}}
\def\IP{\relax{\rm I\kern-.18em P}}
\def\IZ{\relax{\rm I\kern-.18em Z}}

%
%%%%%%%%%%%%%%%%%%%%%%%%%%%%%%%%%%%%
%
\def\np#1#2#3{Nucl. Phys. {\bf B#1} (#2) #3}
\def\pl#1#2#3{Phys. Lett. {\bf #1B} (#2) #3}
\def\plb#1#2#3{Phys. Lett. {\bf #1B} (#2) #3}

\def\prd#1#2#3{Phys. Rev. {\bf D#1} (#2) #3}

\catcode`\@=11
\def\slash#1{\mathord{\mathpalette\c@ncel{#1}}}
\overfullrule=0pt

\def\MM{{\cal M}^{2n}}

\def\underrel#1\over#2{\mathrel{\mathop{\kern\z@#1}\limits_{#2}}}

\catcode`\@=12

%%%%%%%%%%%%%%%%%%%%%%%%%%%%%%%%%%%%%%%%%%%%%%%%%%%%%%%%%%%%%%

%

%%%%%%%%%%%%%%%%%%%%%%%%%%%%%%%%%%%%%%%%%%%%%%%%%%%%%%%%%%%%%%
% new defs:

\rightline{NSF-ITP-99-123}
\rightline{RI-10-99, EFI-99-45}
\Title{
\rightline{hep-th/9911039}}
{\vbox{\centerline{Comments on Double Scaled Little String Theory}}}
\medskip
\centerline{\it Amit Giveon${}^{1,2}$ and David Kutasov${}^{1,3}$}
\bigskip
\centerline{${}^1$Institute for Theoretical Physics}
\centerline{University of California, Santa Barbara, CA 93106, USA}
\smallskip
\centerline{${}^2$Racah Institute of Physics, The Hebrew University}
\centerline{Jerusalem 91904, Israel}
\centerline{giveon@vms.huji.ac.il}
\smallskip
\centerline{${}^3$Department of Physics, University of Chicago}
\centerline{5640 S. Ellis Av., Chicago, IL 60637, USA }
\centerline{kutasov@theory.uchicago.edu}

\lref\gk{A. Giveon and D. Kutasov, hep-th/9909110.}

\bigskip\bigskip\bigskip
\noindent
We study little string theory in a weak coupling limit defined in \gk.

\vfill

\Date{10/99}

\newsec{Introduction}

\lref\natistr{N. Seiberg, hep-th/9705221, \pl{408}{1997}{98}.}%
\lref\dvv{R. Dijkgraaf, E. Verlinde and H. Verlinde, hep-th/9604055,
Nucl. Phys. {\bf B486} (1997) 89; hep-th/9704018, Nucl. Phys. {\bf B506} 
(1997) 121.}%
\lref\brs{M. Berkooz, M. Rozali and N. Seiberg, hep-th/9704089, Phys. Lett.
{\bf 408B} (1997) 105.}%
\lref\OV{H. Ooguri and C. Vafa, hep-th/9511164, \np{463}{1996}{55}.}%
\lref\Kutasov{D. Kutasov, hep-th/9512145, \plb{383}{1996}{48}.}%
\lref\klmvw{A. Klemm, W. Lerche, P. Mayr, C. Vafa and N. Warner,
hep-th/9604034, Nucl. Phys. {\bf B477} (1996) 746.}%
\lref\ghm{R. Gregory, J. Harvey and G. Moore, hep-th/9708086,
Adv. Theor. Math. Phys. {\bf 1} (1997) 283.}%

Little String Theories (LST's) are non-local theories without gravity
which can be defined by studying the dynamics on $NS5$-branes
in the limit in which the string coupling $g_s\to 0$ \natistr\ 
(see also \refs{\dvv,\brs}). An alternative (and possibly more general) 
definition of these theories is obtained by studying string theory on 
\eqn\cym{\IR^{d-1,1}\times \MM~,} 
where $\MM$ is a singular Calabi-Yau (CY) $n$-fold ($2n=10-d$).
In the limit $g_s\to 0$ the modes that propagate in the bulk of $\MM$
decouple and one finds a non-trivial theory living at the singular
locus of $\MM$. In some cases the two definitions are related
by T-duality \refs{\OV,\Kutasov,\klmvw,\ghm}.

\lref\abks{O. Aharony, M. Berkooz, D. Kutasov and N. Seiberg,
hep-th/9808149, JHEP {\bf 9810} (1998) 004.}%
\lref\imsy{N. Itzhaki, J. Maldacena, J. Sonnenschein and S.
Yankielowicz, hep-th/9802042, \prd{58}{1998}{046004}.}%
\lref\bst{H.J. Boonstra, K. Skenderis and P.K. Townsend,
hep-th/9807137, JHEP {\bf 9901} (1999) 003.}%
\lref\dg{D. Diaconescu and J. Gomis, hep-th/9810132,
\np{548}{1999}{258}.}%
\lref\kll{D. Kutasov, F. Larsen and R. Leigh, hep-th/9812027,
\np{550}{1999}{183}.}%
\lref\aaabbb{O. Aharony and T. Banks, hep-th/9812237, 
JHEP {\bf 9903} (1999) 016.} 
\lref\minsei{S. Minwalla and N. Seiberg,  hep-th/9904142,
JHEP {\bf 9906} (1999) 007.}%
\lref\gka{M. Gremm and A. Kapustin, hep-th/9907210.}%
\lref\aabb{O. Aharony and M. Berkooz, hep-th/9909101.}
\lref\gkp{A. Giveon, D. Kutasov and O. Pelc, hep-th/9907178.}

A large class
of examples is obtained by studying CY $n$-folds $\MM$ \cym\ with
an isolated quasi-homogenous hypersurface singularity. Near such a 
singularity, $\MM$ can be described as the hypersurface $F(z_1,\cdots,
z_{n+1})=0$ in $\IC^{n+1}$, where $F$ is a quasi-homogenous polynomial 
with weight one under $z_a\to\lambda^{r_a}z_a$, \ie\
\eqn\homog{F(\lambda^{r_a}z_a)=\lambda F(z_a)~, \;\;\lambda\in \IC~,}
for some set of positive weights $r_a$. 

Finding a useful description of LST's is an important open problem.
In \abks\ it was proposed to study them using holography (see also 
\refs{\imsy,\bst,\dg,\kll,\aaabbb,\minsei,\gka,\aabb}). In \gkp\ it was further 
argued that the non-gravitational $d$-dimensional theory at a general 
singularity $F(z_a)=0$ is dual to string theory in the background
\eqn\rdd{\IR^{d-1,1}\times\IR_\phi\times S^1\times LG(W=F)~,}
where $\IR_\phi$ is the real line, labeled by $\phi$, along which
the dilaton varies linearly,
\eqn\lindil{\Phi=-{Q\over2}\phi~,}
and $LG(W=F)$ is a Landau-Ginzburg $N=2$ SCFT of $n+1$ chiral
superfields $z_a$ with superpotential $W(z_a)=F(z_a)$ \homog.
{\it Off-shell} physical observables in LST correspond to 
{\it on-shell} observables in string theory on \rdd. 
Off-shell Green's functions in LST correspond to
on-shell correlation functions in the dual string theory.

The description \rdd\ is useful at large $\phi$ where the string
coupling $e^\Phi\to 0$ and the theory is weakly coupled. 
One can use it to identify off-shell observables in LST (which
correspond in the dual description to wavefunctions which are 
exponentially supported at $\phi\to\infty$) and study their transformation
properties under the symmetries of the theory \abks. Correlation functions
are governed by the region of finite $\phi$, and therefore computing
them requires knowledge of the strong coupling dynamics at $\phi\to-
\infty$. Perturbative string theory in the vacuum \rdd\ is singular
and does not provide a good guide for such correlation functions.

The strong coupling singularity at $\phi\to-\infty$ in \rdd, \lindil\ can be avoided 
by replacing the cylinder $\IR_\phi\times S^1$ in \rdd\ by the SCFT on the 
``cigar'' $SL(2)/U(1)$ (or equivalently \gk\ $N=2$ Liouville). In the dual theory
this corresponds \refs{\gkp,\gk} to resolving the singularity $F(z_a)=0$ to
\eqn\noncomp{F(z_1,\cdots, z_{n+1})=\mu~,}
and taking the limit $\mu\to 0$, $g_s\to 0$ with the ratio
\eqn\scvar{\chi\equiv {\mu^{r_\Omega}\over g_s}}
held fixed. Here
\eqn\rom{r_\Omega\equiv\sum_{a=1}^{n+1} r_a-1={Q^2\over 2}}
(see \refs{\gkp,\gk} for the details). 
The value of the string coupling at the tip of the cigar is
$1/\chi$; hence, by making $\chi$ large one obtains a weakly coupled
theory, Double Scaled Little String Theory (DSLST), in which 
correlation functions have a $1/\chi$ expansion. 

\lref\sfet{K. Sfetsos, hep-th/9811167, JHEP {\bf 9901} (1999) 015;
hep-th/9903201.} %

To summarize, the non-gravitational $d$-dimensional theory
which lives at the resolved singularity \noncomp\ in the double scaling
(decoupling) limit $\mu, g_s\to 0$ with $\chi$ \scvar\ held fixed is dual to string
theory in the background 
\eqn\fullst{\IR^{d-1,1}\times{SL(2)_k\over U(1)}\times LG(W=F)~,}
where $k$ is the level of $SL(2)$ (it is related to \rom, $k=1/r_\Omega$ 
\gkp). One can think of \fullst\ as the ``near horizon geometry'' of the
resolved singularity $F=\mu$ \noncomp\ \sfet. 

\lref\gkreview{A. Giveon and D. Kutasov, hep-th/9802067, 
Rev. Mod. Phys. {\bf 71} (1999) 983.}% 

An interesting example of the construction is the $d=6$
theory of $k$ parallel $NS5$-branes. In that case one has
\eqn\fott{F(z_1,z_2,z_3)=z_1^k+z_2^2+z_3^2~,}
and the manifold $F(z_a)=0$ is an ALE space (the singular
geometry which is T-dual to the fivebrane near horizon background). 
The low energy limit
of the theory on $k$ $NS5$-branes in type IIB string theory is
$SU(k)$ gauge theory with $(1,1)$ SUSY (a non-chiral theory with
sixteen supercharges);
for type IIA fivebranes one finds at low energies a mysterious
non-Abelian generalization of the theory of self-dual $B_{\mu\nu}$
fields, with chiral $(2,0)$ SUSY  and conformal invariance (see \eg\
\gkreview\ for a review).
At high energies both the IIA and IIB theories are expected
to have a Hagedorn density of states, but unlike ordinary critical
string theories they do not contain gravity and
do have well defined off-shell Green's functions.

\lref\chs{C. Callan, J. Harvey and A. Strominger, hep-th/9112030,
in Trieste 1991, Proceedings, String Theory and Quantum Gravity 1991,
208.}

The dual background
\rdd\ is in this case $\IR^{5,1}\times \IR_\phi\times S^1\times
SU(2)_k/U(1)$. The GSO projection acts as a $Z_k$ orbifold on
$S^1\times SU(2)_k/U(1)$, turning it into $SU(2)_k$ WZW, the CFT
on a three-sphere. This theory has an $SU(2)_L\times SU(2)_R$ symmetry
which in the fivebrane language is the $SO(4)$ $R$-symmetry of rotations
in the $\IR^4$ transverse to the branes \chs.  

Turning on $\mu$ \noncomp\ corresponds in the fivebrane language to
distributing the fivebranes at equal distances around a circle of radius $r_0\sim
l_s\mu^{1\over k}$ (see \gk\ for details). Parametrizing the $\IR^4$ transverse 
to the branes by two complex variables $A$, $B$, the fivebrane theory
includes two $k\times k$ matrices of complex scalar fields which we
also denote by $A$, $B$. These scalars belong to the same
SUSY multiplets as the $SU(k)$ gauge field in type IIB, and the non-Abelian
self-dual $B_{\mu\nu}$ field and an additional real adjoint scalar in the IIA theory.

The double scaling limit \noncomp\ - \rom\ corresponds to studying the 
theory at a point in its Coulomb branch where, say,
\eqn\bbb{\eqalign{\langle A\rangle=&0\cr
\langle B\rangle=& M_WM_s{\rm diag} (e^{2\pi i\over k}, e^{4\pi i\over k},
\cdots, e^{2\pi ik\over k})~.\cr}}
In type IIB string theory, this Higgses the $SU(k)$ gauge symmetry down
to $U(1)^{k-1}$ with the mass of the charged gauge bosons 
$M_{jl}\sim M_W|e^{2\pi i j\over k}-e^{2\pi i l\over k}|$.
These gauge bosons correspond to the ground states of D-strings stretched
between the $NS5$-branes; thus, $M_W$ is related to $r_0$ via
\eqn\mwro{M_W\sim {r_0\over g_s l_s^2}\sim \chi M_s~.}
Therefore, the double scaling limit is in this case the decoupling 
limit with the mass of the $W$-bosons held fixed. 

In the IIA theory we have instead $D2$-branes stretched between the $NS5$-branes;
these look like strings in the fivebranes, charged under the respective self-dual
$B_{\mu\nu}$ fields on the fivebranes. The tensions of these strings are proportional
to $M_{jl}M_s$. They are held finite in the double scaling limit. 

Distributing the branes as in \bbb\ breaks the $R$-symmetry
\eqn\breakR{SO(4)\to SO(2)\times Z_k~.} 
The charges of $A$ and $B$ under these symmetries
are $(1,0)$ and $(0,1)$, respectively. In the dual picture \fullst, 
the $SO(2)$ and $Z_k$ charges correspond to momentum and winding around
the cigar, respectively\foot{In the language of \gkp, the $SO(2)$ and $Z_k$
charges are ${1\over2}(R-\bar R)$ and ${1\over2}(R+\bar R)\;{\rm mod}\; k$, 
respectively.} (recall that in the cigar CFT, momentum around the cigar is 
conserved, while winding number is not).

\newsec{Observables in DSLST}

\lref\teschner{J. Teschner, hep-th/9712256; hep-th/9712258;
hep-th/9906215; V.A. Fateev, A.B. Zamolodchikov and Al.B. 
Zamolodchikov, unpublished.}%

We are interested in constructing BRST invariant observables in string
theory on \fullst.
We will focus on (NS,NS) sector fields, since the rest of the observables
can be obtained from these by applying the spacetime supercharges \gkp.
Such observables are linear combinations of vertex operators of the form
(see \refs{\gk,\gkp} for details)
\eqn\mostgenobs{O_{W,m,\bar m}(k_\mu)=e^{-\varphi-\bar\varphi}
W_{\Delta,\bar\Delta} e^{ik_\mu x^\mu}V_{j;m,\bar m}~,}
where $k_\mu$ $(\mu=0,1,\cdots, d-1)$ is the momentum along
$\IR^{d-1,1}$  and $W_{\Delta,\bar\Delta}$ is a rather
general operator\foot{By a gauge choice it can be set to be the 
identity on the cigar, and it will be convenient to do that below.} 
with scaling dimension $(\Delta,\bar\Delta)$ in the SCFT on 
\fullst. $V_{j;m,\bar m}$ is a Virasoro primary on $SL(2)/U(1)$. 
Its scaling dimensions are
\eqn\scdimv{\Delta_{j;m,\bar m}={m^2-j(j+1)\over k}~;\;\;
\bar\Delta_{j;m,\bar m}={\bar m^2-j(j+1)\over k}~,}
where $(m,\bar m)$ run over the set 
\eqn\mbarm{m={1\over2}(p+wk)~;\;\;\bar m=-{1\over2}(p-wk)~.}
One can think of $p$ as the momentum number around the cigar (which
is conserved) and $w$ as the winding number (which is not). In CFT on
$SL(2)/U(1)$ both $p$ and $w$ are integers. String theory on \fullst\ 
further includes a chiral orbifold (the GSO projection) acting non-trivially
on $SL(2)_k/U(1)\times LG(W=F)$ in \fullst. The twisted sectors contain
in general operators with non-integer winding number $w$ which is conserved
modulo one; $p$ remains integer.

\lref\morerefs{
J. Balog, L. O'Raifeartaigh, P. Forgacs, and A. Wipf,
Nucl. Phys. {\bf B325} (1989) 225;
L.J. Dixon, M.E. Peskin and J. Lykken,  Nucl. Phys. {\bf B325} (1989) 329;
P.M.S. Petropoulos, Phys. Lett. {\bf B236} (1990) 151;
I. Bars and D. Nemeschansky, Nucl. Phys. {\bf B348} (1991) 89;
S. Hwang, Nucl. Phys. {\bf B354} (1991) 100;
K. Gawedzki, hep-th/9110076;
I. Bars, Phys. Rev. {\bf D53} (1996) 3308, hep-th/9503205;
in {\it Future Perspectives In String Theory} (Los Angeles, 1995),
hep-th/9511187;
J.M. Evans, M.R. Gaberdiel, and M.J. Perry, hep-th/9812252.}%

The on-shell condition requires (the coefficient of
$e^{-\varphi-\bar\varphi}$ in) \mostgenobs\ to be a bottom 
component of a worldsheet $N=1$ superfield, and in addition to satisfy
(we will usually set $M_s=1$ from now on)
\eqn\physstgen{\eqalign{
{1\over2}k_\mu k^\mu+{m^2-j(j+1)\over k}+\Delta=&{1\over2}~,\cr
{1\over2}k_\mu k^\mu+{\bar m^2-j(j+1)\over k}+\bar\Delta=&{1\over2}~.\cr
}}
The chiral GSO projection further requires
\eqn\gsop{Q_{LG}+{2m\over k}+F_L\in 2Z+1~,}
where $Q_{LG}$ is the charge of $W_{\Delta,\bar\Delta}$ under the $U(1)_R$
symmetry which belongs to the left-moving $N=2$ superconformal algebra in 
$LG(W=F)$, and $F_L$ is the left-moving fermion number on $\IR^{d-1,1}$.
A similar relation holds for the right-movers. Unitarity of the ``near horizon 
CFT,'' which is an orbifold of $SL(2)/U(1)\times LG(W=F)$, implies 
that $j$ belongs to one of the following two regions (for early work
on unitarity of $SL(2)/U(1)$ see \morerefs):
\eqn\regone{\eqalign{
&j\in \IR;\;\;\;-{1\over2}<j<{k-1\over2}\cr
&{m^2-j(j+1)\over k}+\Delta_{LG}\ge 0\cr
&{\bar m^2-j(j+1)\over k}+\bar\Delta_{LG}\ge 0\cr
}}
or
\eqn\regtwo{j\in-{1\over2}+i\IR~,}
where $\Delta_{LG}$ is the contribution to $\Delta$ of the LG model in \fullst.
Off-shell observables in the $d$-dimensional LST correspond to wave-functions
on the cigar that are exponentially supported far from the tip, at
$\phi\to\infty$. Therefore, only vertex operators in the range
\regone\ give rise to such observables. Operators in the range 
\regtwo\ are $\delta$ function normalizable, and correspond
to fluctuating fields in LST.

Note that the bound on $j$ \regone\ implies via \physstgen\ that
off-shell physical observables \mostgenobs\ are only defined in
a finite range of values of $k_\mu k^\mu$. This issue will be further
discussed below.

It is instructive to apply the above discussion to the six dimensional
theory of $k$ $NS5$-branes in type II string theory. The string
vacuum 
\eqn\stvcsix{\IR^{5,1}\times {SL(2)_k\over U(1)}\times {SU(2)_k\over U(1)}}
has the same $R$-symmetry, $SO(2)\times Z_k$, as the brane theory \breakR.
The $SO(2)$ charge is $m-\bar m=p$ \mbarm, while the $Z_k$ charge
is $(m+\bar m) \, {\rm mod}\, k$. 

In the fivebrane theory there are two sets of chiral operators\foot{The $k\times k$ 
matrix scalar fields $A$ and $B$ were defined before equation \bbb. In DSLST the 
off-diagonal components of $A$, $B$ are massive; the eigenvalues of the matrices 
are the light degrees of freedom.}: 
\eqn\aabb{{\rm Tr}\,A^i~, \qquad {\rm Tr}\, B^i~, \qquad i=2,\cdots, k~.}
The $SO(2)\times Z_k$ charge of $A^i$ is $(i,0)$; that of $B^i$ is
$(0,i)$. In the string vacuum \stvcsix\ they are realized as follows:
\eqn\BBB{\eqalign{
&{\rm Tr} B^{2m}\leftrightarrow e^{-\varphi-\bar\varphi}z_1^{k-2m}
e^{ik_\mu x^\mu}V_{j;m,m}\cr
&2m=2,3,\cdots, k\cr
&k_\mu^2={2\over k}(j-m+1)(j+m)\cr
}}
\eqn\AAA{\eqalign{
&{\rm Tr} A^{2m}\leftrightarrow e^{-\varphi-\bar\varphi}\tilde z_1^{k-2m}
e^{ik_\mu x^\mu}V_{j;m,-m}\cr
&2m=2,3,\cdots, k\cr
&k_\mu^2={2\over k}(j-m+1)(j+m)~,\cr
}}
where $z_1^i$ are the chiral-chiral operators in the $N=2$ minimal model
$SU(2)/U(1)$ (see \fott), and $\tilde z_1^i$ are their chiral-antichiral analogs,
which appear in the twisted sectors of the GSO orbifold. The following comments
regarding the correspondence \BBB, \AAA\ are in order at this point:
\item{(1)} The $SO(2)\times Z_k$ (see \breakR) charges of the vertex operators
on the r.h.s. of \BBB, \AAA\ are $(0,2m)$ and $(2m,0)$, respectively, in agreement
with those of ${\rm Tr} B^{2m}$ and ${\rm Tr} A^{2m}$.
\item{(2)} The supersymmetry multiplet structure is the same. In particular,
the vertex operators in \BBB, \AAA\ preserve the same halves of the SUSY 
as the operators on the l.h.s. \gkp.
\item{(3)}  Far from the tip of the cigar, where it looks like a cylinder, one recovers
the background $\IR_\phi\times S^3$ studied in \abks. Roughly speaking,
large $\phi$ corresponds to high energies on the fivebrane, a regime in 
which the Higgs breaking due to \bbb\ is negligible. The vertex 
operators\foot{Which can be thought of as ``tachyons'' from  the $d$-dimensional 
perspective \refs{\gkp,\gk}.} in \BBB, \AAA\
approach for large $\phi$ the ``graviton'' operators given by eq. (3.6) 
in \abks. The identification \BBB, \AAA\ agrees with the
one proposed in \abks\ in that limit. 
\item{(4)} There are analogs of the vertex operators in \BBB, \AAA\
with $m\not=\bar m$. These correspond to ${\rm Tr} A^{m-\bar m} B^{m+\bar m}$.
Such operators do not preserve any spacetime SUSY.

\newsec{The two point functions}

\lref\gko{P. Goddard, A. Kent and D. Olive, Phys. Lett.{\bf 152B} (1985) 88.}%
\lref\fatzam{A.B. Zamolodchikov and V.A. Fateev, Sov. J. Nucl. Phys. {\bf 43}
(1986) 657.}%

The CFT on the cigar is related by the GKO coset construction \gko\ to that
on $AdS_3$. The natural observables in CFT on the Euclidean version of $AdS_3$ 
(the manifold $H_3^+=SL(2,C)/SU(2)$) are functions $\Phi_j(x,\bar x;z,\bar z)$
which transform as primaries under the $SL(2)_L\times SL(2)_R$ current algebra 
\refs{\fatzam, \teschner}.
$x$ is an auxiliary complex variable that is useful for studying CFT on $AdS_3$.
The two point function of $\Phi_j$ is\foot{Here and below we suppress the
dependence of correlation functions on the worldsheet locations $z$, $\bar z$.}
\eqn\twoptpj{\langle \Phi_j(x,\bar x)\Phi_j(x',\bar x')\rangle=
{k\over\pi}[\nu(k)]^{2j+1}{\Gamma(1-{2j+1\over k})\over \Gamma({2j+1\over k})}
|x-x'|^{-4(j+1)}~,}
where
\eqn\nuk{\nu(k)\equiv{1\over\pi} {\Gamma(1+{1\over k})\over \Gamma(1-{1\over k})}~.}
For studying the coset it is convenient to ``Fourier transform''
the fields $\Phi_j(x,\bar x)$ and define the mode operators
\eqn\modeops{\Phi_{j;m,\bar m}=\int d^2x\, x^{j+m}\bar x^{j+\bar m}
\Phi_j(x,\bar x)~.}
Note that for \modeops\ to make sense one needs $m-\bar m\in Z$;
as discussed above (see \mbarm) this is indeed always the case. 

The two point functions of the modes of $\Phi$ are equal to those of the 
$SL(2)/U(1)$ coset theory,
\eqn\vvpp{\langle V_{j;m,\bar m} V_{j; -m,-\bar m}\rangle=
\langle\Phi_{j;m,\bar m} \Phi_{j; -m,-\bar m}\rangle}
(the two differ by the two point function of exponentials
in CFT on $S^1$, which is equal to one\foot{$V_{j;m,\bar m}$ defined by 
\modeops, \vvpp\ is normalized such that far from the tip of the cigar
it describes an incoming wave with strength one (see eq. (3.10) in 
\gk).}).

The two point function of \mostgenobs\ is a generalization of
the one computed in \gk\ for $m=\bar m$ using the results of \teschner:
\eqn\twopt{G_2(k_\mu)\equiv\langle O_{W,m,\bar m}(k_\mu)
O_{W',-m,-\bar m}(-k_\mu)\rangle=
\langle W_{\Delta,\bar\Delta} W'_{\Delta,\bar\Delta}\rangle\langle 
V_{j;m,\bar m}V_{j;-m,-\bar m}
\rangle~.}
Using \vvpp, \modeops, \twoptpj\ and the results of Appendix A one finds
\eqn\twoptvjm{\langle V_{j;m,\bar m}V_{j;-m,-\bar m}\rangle=k[\nu(k)]^{2j+1}
{\Gamma(1-{2j+1\over k})\Gamma(-2j-1)\Gamma(j-m+1)\Gamma(1+j+\bar m)
\over
\Gamma({2j+1\over k})\Gamma(2j+2)\Gamma(-j-m)\Gamma(\bar m-j)}~.}
The two point function \twoptvjm\ has a series of poles; these can be
interpreted as contributions of on-shell states in DSLST, which are
created from the vacuum by the operator $O_{W,m,\bar m}$. The masses of these
states can be computed by using the relation \physstgen\ between $j$ and 
$M^2=-k_\mu k^\mu$. The resulting
spectrum of masses can be analyzed as in \gk. 
Restricting to the region \regone\ gives rise to non-negative values
of $M^2$ corresponding to 
\eqn\poleone{m=j+n;\;\;\;n=1,2,3,\cdots}
or
\eqn\poletwo{\bar m=-(j+n);\;\;\;n=1,2,3,\cdots}
These values of $m$ and $j$ belong to the principal discrete series
representations of $SL(2)$. The corresponding states can be thought
of as bound states that live near the tip of the cigar \ref\dvvo{R. 
Dijkgraaf, E. Verlinde and H. Verlinde, Nucl. Phys. {\bf B371} (1992) 
269.}. Without loss of generality, we will restrict to the set \poleone.

Near $m=j+n$ the two point function \twopt\ takes the form 
\eqn\twoptpole{G_2={\langle W_{\Delta,\bar\Delta }
W'_{\Delta,\bar\Delta}\rangle\over j+n-m}k[\nu(k)]^{2j+1}
{\Gamma(1-{2j+1\over k})\Gamma(1+j+\bar m)\over \Gamma({2j+1\over k})\Gamma(2j+2)
\Gamma(\bar m-j)}
{1\over (n-1)!}\prod_{i=2}^n(2j+i)+\cdots}
where the ``$\cdots$'' correspond to terms analytic near $m=j+n$.
Using \physstgen, one can rewrite \twoptpole\ as
\eqn\mompole{ G_2(k_\mu)={R_{n,m}\over k_\mu^2+M^2_{n,m}}+\cdots}
where \gk
\eqn\kmuM{{1\over k_\mu^2+M_{n,m}^2}={k\over 2(2j+1)(j+n-m)}~,}
\eqn\massgen{\eqalign{
&M_{n,m}^2=2(\Delta+{m\over k})-1+{2\over k}(n-1)(2m-n)~,\cr
&2m+1>2n>2m-k+1~,\cr}}
\eqn\RRR{R_{n,m}=2\langle W_{\Delta,\bar\Delta }
W'_{\Delta,\bar\Delta}\rangle[\nu(k)]^{2j+1}
{\Gamma(1-{2j+1\over k})\Gamma(1+j+\bar m)\over \Gamma({2j+1\over k})
\Gamma(2j+2)\Gamma(\bar m-j)}
{1\over (n-1)!}\prod_{i=1}^n(2j+i)~.}
It is useful to note that:
\item{(1)} Since $m-\bar m=p$ \mbarm, we have $\bar m=j+n-p$ (recall that $n,p\in Z$).
\item{(2)} When $|\bar m|\le j$, the residue \RRR\ vanishes due to
a pole of $\Gamma(\bar m-j)$. This is in agreement with the fact
that, as explained in \gk, physical particles correspond to the
principal discrete series representations (for which $|\bar m|\ge j+1$).
\item{(3)} If $n>p$, \ie\ $\bar m=j+\bar n$, $\bar n=1,2,\cdots$,
the residue \RRR\ is positive for all $\bar n$, in agreement with the
expected unitarity of the DSLST.
\item{(4)} For $n\le p$, comment (2) above implies that non-zero
residues correspond to $p-n>2j$ and, furthermore, one must have
$2j\in Z$, in which case the pole of $\Gamma(\bar m-j)$ in the 
denominator is cancelled by a pole of $\Gamma(1+j+\bar m)$ in the
numerator. The sign of the residue \RRR\ does not vary with $n$. 
It does depend on the observable under consideration as $(-)^{2m+1}$,
however, this dependence is insignificant. For example, one can 
absorb it into the rules for Hermitian conjugation\foot{A similar phenomenon
occurs in $SU(2)$  WZW, where the $SU(2)_L\times SU(2)_R$ symmetry implies that 
the two point functions satisfy the relation
$\langle V_{j;m,\bar m}V_{j;-m,-\bar m}\rangle= (-)^{2m}
\langle V_{j;m,-\bar m}V_{j;-m,\bar m}\rangle$. Unitarity of the CFT
then implies that one should use conjugation rules that are similar
to the ones that appear here. Another potential source of minus 
signs is the relation between observables on $SL(2)$ and $SL(2)/U(1)$, \vvpp.}:
\eqn\hermcon{V_{j;m,\bar m}^\dagger=\cases{V_{j;-m,-\bar m}& $m\bar m>0$\cr
(-)^{2m+1}V_{j;-m,-\bar m}& $m\bar m<0$\cr
}}

\noindent
To summarize, the two point function \twopt\ exhibits a series of
single poles, all of whose residues have the same sign (for a given
observable). This is consistent with the expected unitarity of the 
$d$-dimensional theory.

In addition to the series of poles described above, the two point function
\twopt\ has a branch cut in momentum space, since
$j$ is determined in terms of $k_\mu k^\mu$ by solving the quadratic equation
\physstgen. This branch cut starts at $j=-1/2$ and is due to the fact that
the operators $O_{W,m, \bar m}$ can create from the vacuum the continuum of normalizable 
states in the range \regtwo. However, as discussed above, non-fluctuating
physical observables correspond only to operators in the range \regone\
and, therefore, the role of both the branch cut and the observables
\regtwo\ is not clear.

The analysis of the mass spectrum provides another check of the
identification of the chiral operators \aabb\ in the six dimensional
fivebrane theory, at the point \bbb\ in its Coulomb branch, with the vertex
operators \BBB, \AAA. The two point functions of these operators have
poles given by \massgen\ with $\Delta(z_1^{k-2m})=\Delta(\tilde z_1^{k-2m})
=(k-2m)/2k$. The lowest lying states created by these operators from the vacuum
correspond to $n=1$ in \massgen, and are massless.  Thus, the operators
\BBB, \AAA\ create from the vacuum $4(k-1)$ real scalar fields with the
quantum numbers of the massless scalars corresponding to \aabb\ in the 
fivebrane theory.

There is however a puzzle associated with the above discussion.
At low energies, the eigenvalues of the matrices $A$, $B$ are expected
to be free massless fields. In free field theory, operators like \aabb\ 
do not create single particle states from the vacuum: ${\rm Tr}\, A^i$ 
(say) creates an $i$-particle state. Put differently, in free field theory
the two point function $\langle {\rm Tr}\, A^i(k_\mu) {\rm Tr}\, A^i(-k_\mu)\rangle$ 
behaves at small momenta like $(k_\mu k^\mu)^{2i-3}\log k_\mu k^\mu$ and {\it not} 
$1/k_\mu^2$ as found in \twopt\ -- \RRR. We will discuss this puzzle further in 
section 5.

\newsec{The three point functions}

To compute the couplings among the particle states with masses given by
\massgen\ we next turn to the three point functions of the off-shell
observables \mostgenobs, 
\eqn\threee{G_3\equiv\langle O_{W_3,m_3,\bar m_3}O_{W_2,m_2,\bar m_2}
O_{W_1,m_1,\bar m_1}\rangle~.}
The momentum around the cigar, $m-\bar m$, is conserved;
hence three point functions satisfy
\eqn\consvmom{m_1+m_2+m_3=\bar m_1+\bar m_2+\bar m_3~.}
As is standard in fermionic string theory, 
to compute such three point functions we take two of the operators to
be in the $-1$ picture, and the third in the $0$ picture. After evaluating
the ghost contributions, one finds:
\eqn\threept{
G_3=\langle W_{\Delta_3,\bar\Delta_3}e^{ik_3\cdot x} V_{j_3;m_3,\bar m_3}
W_{\Delta_2,\bar\Delta_2}e^{ik_2\cdot x} V_{j_2;m_2,\bar m_2}
G_{-{1\over2}}\bar G_{-{1\over2}} W_{\Delta_1,\bar\Delta_1}e^{ik_1\cdot x} 
V_{j_1;m_1,\bar m_1}\rangle~.}
$G_{-{1\over2}}$ and $\bar G_{-{1\over2}}$ are the left and right-moving
global $N=1$ superconformal generators. 

There are apriori four contributions to \threept: each of $G_{-1/2}$ and 
$\bar G_{-{1\over2}}$ can act either on $W_1 e^{ik_1\cdot x}$ or on $V_{j_1;m_1,\bar m_1}$.
The ``mixed terms'' in which the two supercharges act in different sectors
vanish due to \consvmom\ and the left and right-moving $U(1)_R$ $N=2$
superconformal symmetries. This leaves two contributions: 
\eqn\threere{\eqalign{G_3=&\langle W_{\Delta_3,\bar\Delta_3}e^{ik_3\cdot x}
                         W_{\Delta_2,\bar\Delta_2}e^{ik_2\cdot x}
G_{-{1\over2}}\bar G_{-{1\over2}} W_{\Delta_1,\bar\Delta_1}e^{ik_1\cdot x}\rangle
\langle V_{j_3;m_3,\bar m_3}V_{j_2;m_2,\bar m_2}V_{j_1;m_1,\bar m_1}\rangle\cr
+&\langle W_{\Delta_3,\bar\Delta_3}e^{ik_3\cdot x}
W_{\Delta_2,\bar\Delta_2}e^{ik_2\cdot x}
W_{\Delta_1,\bar\Delta_1}e^{ik_1\cdot x}\rangle
\langle V_{j_3;m_3,\bar m_3}V_{j_2;m_2,\bar m_2}
G_{-{1\over2}}\bar G_{-{1\over2}}V_{j_1;m_1,\bar m_1}\rangle~.\cr
}}
The $U(1)_R$ symmetries imply that the first term vanishes unless
$m_1+m_2+m_3=0$. Thus, it corresponds to amplitudes
for which both momentum \consvmom\ and winding are conserved.

The second term in \threere\ is in general non-zero when
$m_1+m_2+m_3=\pm{k\over2}$. The $N=1$ superconformal generator
$G_{-{1\over2}}$ can be decomposed into eigenstates of $U(1)_R$,
$G_{-{1\over2}}=G_{-{1\over2}}^++G_{-{1\over2}}^-$. The $U(1)_R$ 
constraints imply that in the second term in \threere\  one has to 
act on $V$ either with $G_{-{1\over2}}^+\bar G_{-{1\over2}}^+$ or
with $G_{-{1\over2}}^-\bar G_{-{1\over2}}^-$. The resulting amplitudes 
violate winding number by one unit. It is interesting to note that:
\item{(1)} One can generalize the above arguments
to $n$ point functions. The correlation
function receives contributions that violate winding number
by $i$ units, with $i=0,1,\cdots, n-2$. 
\item{(2)} A related result was obtained by V. Fateev, A.B. Zamolodchikov 
and Al.B. Zamolodchikov (unpublished), who showed that the bosonic CFT on 
$SL(2)/U(1)$ has similar properties: $n$ point functions can violate winding 
number conservation by up to $n-2$ units.

\noindent
We next turn to the calculation of the three point function \threere.
For simplicity, we will present the result for the case where all
three momenta $m_i-\bar m_i$ vanish, and the winding number is conserved.
In that case, only the first line of \threere\ contributes. Furthermore, 
the analytic structure is determined by the three point function on
the cigar, 
\eqn\threesimp{
\langle V_{j_3;m_3, m_3}V_{j_2;m_2, m_2}V_{j_1;m_1, m_1}\rangle~,}
and we will focus on it below.

In the interpretation of the correlation function $G_3$ as an off-shell
Green's function in DSLST, one expects \threere, \threesimp\ to have
poles at values of $m$ and $j$ corresponding to the principal discrete
series \poleone, \poletwo. Such poles would be due to the states found in
section 3 (see \massgen) going on mass shell. As in local QFT, 
LSZ reduction relates the residue of these poles to the S-matrix element
describing (say) the decay of one of these particles to two others. Below
we will compute these matrix elements.

The three point function \threesimp\ is obtained from the appropriate 
correlation function of $SL(2)_L\times SL(2)_R$ primaries in CFT on $H_3^+$,
using the transform \modeops. As shown in \teschner,
\eqn\thpf{\eqalign{
&\langle \Phi_{j_3}(x_3,\bar x_3)\Phi_{j_2}(x_2,\bar x_2)
\Phi_{j_1}(x_1,\bar x_1)\rangle=\cr
&D(j_3,j_2,j_1;k)|x_1-x_2|^{2(j_3-j_1-j_2-1)}
|x_1-x_3|^{2(j_2-j_1-j_3-1)}|x_2-x_3|^{2(j_1-j_2-j_3-1)}~,\cr}}
where 
\eqn\djk{\eqalign{
&D(j_3,j_2,j_1;k)={k\over 2\pi^3}\nu(k)^{j_1+j_2+j_3+1}\times\cr
&{G(-j_1-j_2-j_3-2)G(j_3-j_1-j_2-1)G(j_2-j_1-j_3-1)
G(j_1-j_2-j_3-1)\over G(-1)G(-2j_1-1)G(-2j_2-1)G(-2j_3-1)}~.\cr}}
$G(j)$ is a special function, whose only properties that will be needed
here are (see also Appendix A):
\item{(1)} It satisfies the recursion relation
\eqn\recrel{G(j-1)={\Gamma(1+{j\over k})\over\Gamma(-{j\over k})} G(j)~.}
\item{(2)} $G(j)$ has poles at the following values of $j$:
$j=n+mk$, $j=-(n+1)-(m+1)k$, $n,m=0,1,2,\cdots$. 

\noindent
A potentially alarming aspect of \thpf, \djk\ is the appearance of
singularities in the three point couplings in string theory on $AdS_3$.
One expects the spacetime theory to be a unitary CFT in which such
singularities are unacceptable.  Upon a closer look one actually
finds that some singularities are acceptable. Consider, for
example, the factor $G(j_1-j_2-j_3-1)$ in the numerator of \djk. 
It has poles when $j_1-j_2-j_3-1=n+mk$ or $-(n+1)-(m+1)k$ with
$n,m=0,1,2,\cdots$. The bound \regone\ satisfied by $j_1$, $j_2$ and $j_3$
eliminates most of these poles and allows only those with 
\eqn\goodpole{j_1-j_2-j_3-1=n=0,1,2,\cdots} 
These poles are harmless and even to some extent necessary. Note
that when \goodpole\ is valid, the power of $|x_2-x_3|^2$ in \thpf\
is a non-negative integer, $n$. Unless the structure constant
$D(j_1, j_2, j_3;k)$ develops a pole in this case, the correlation function
\thpf\ is supported at zero momentum, since it is the Fourier transform of
a polynomial in $x_2-x_3$. The pole at \goodpole\ replaces
\eqn\repxxx{{1\over\epsilon}|x_2-x_3|^{2(n+\epsilon)}\to 
|x_2-x_3|^{2n}\log|x_2-x_3|^2}
whose Fourier transform is a power of momentum, as required. 
The poles \goodpole\ will play an important role below.

Poles of \djk\ that cannot be interpreted as above (\ie\ poles which persist
after \thpf\ is Fourier transformed to momentum space) are genuine
pathologies, and must be avoided. An example of such poles appeared
already in the two point function \twoptpj\ for $j\ge (k-1)/2$; those
poles were excluded by the unitarity bound \regone\ (in fact, excluding
these poles is one way of deriving the bound \gk). 

The three point function \thpf\ has such poles when $j_1+j_2+j_3=n+k-1$, 
$n=0,1,2,\cdots$. These poles have to be excluded by unitarity as well. 
Thus, we conclude that unitarity of weakly coupled string theory on $AdS_3$ 
allows one to consider only three point functions \thpf\ with\foot{This 
constraint on the observables in three point functions may seem
puzzling. We will return to it in the next section.}  
\eqn\badpoles{j_1+j_2+j_3<k-1~.}
We next use the transform \modeops\ and the fact that the three point function
satisfies
\eqn\threevv{\langle V_{j_3;m_3,\bar m_3}V_{j_2;m_2,\bar m_2}V_{j_1;m_1,\bar m_1}
\rangle=\langle \Phi_{j_3;m_3,\bar m_3}\Phi_{j_2;m_2,\bar m_2}\Phi_{j_1;m_1,\bar m_1}
\rangle~,}
like the two point function\foot{The situation for $n\ge 4$ point is more complicated.}
\vvpp. We find
\eqn\basthree{\eqalign{
&\langle V_{j_3;m_3, m_3}V_{j_2;m_2,m_2}V_{j_1;m_1,m_1}
\rangle=D(j_3,j_2,j_1;k)\times\cr
&F(j_3,m_3;j_2,m_2;j_1,m_1)\int d^2 x|x|^{2(m_1+m_2+m_3-1)}~,\cr}}
where
\eqn\jmjmjm{\eqalign{
&F(j_3,m_3;j_2,m_2;j_1,m_1)=\int d^2 x_1 d^2x_2|x_1|^{2(j_1+m_1)}
|x_2|^{2(j_2+m_2)}\times\cr
&|1-x_1|^{2(j_2-j_1-j_3-1)}|1-x_2|^{2(j_1-j_2-j_3-1)}
|x_1-x_2|^{2(j_3-j_1-j_2-1)}~.\cr}}
The integral over $x$ in \basthree\ ensures winding number conservation
$m_1+m_2+m_3=0$. The function $F$ \jmjmjm\ does not seem to be expressible
in terms of elementary functions. 

To study the decay of on-shell
states we must perform LSZ reduction of the three point function, \ie\ 
evaluate $F$ in the vicinity of its singularities. Specifically, since we are
interested (say) in the decay process $1\to 2+3$, we focus on the residue of
the first order pole at 
\eqn\inone{m_1=-(j_1+n_1)~,\qquad n_1=1,2,3,\cdots} 
which arises from the integration over $x_1$ near zero in \jmjmjm. This 
corresponds to an incoming particle with mass determined by \massgen\ 
(compare to \poletwo). One can show that the residue of the pole at
\inone\ has further singularities in the other external legs. In particular,
one finds single poles in $j_2$, $j_3$ at
\eqn\outtwothree{m_2=j_2+n_2~;\;\;\;m_3=j_3+n_3~,\qquad n_2,n_3=1,2,3,\cdots}
corresponding to the outgoing particles $2$ and $3$. Thus, near  \inone, 
\outtwothree, the correlator \threevv\ has the structure
\eqn\resthree{
\langle V_{j_3;m_3, m_3}V_{j_2;m_2, m_2}V_{j_1;m_1, m_1}\rangle
={R_{3,2,1}\over (j_1+n_1+m_1)(j_2+n_2-m_2)(j_3+n_3-m_3)}+\cdots}
We will next compute the residue $R_{3,2,1}$.
Winding number conservation and \inone, \outtwothree\ imply that
\eqn\jojtjt{j_1-j_2-j_3=n_2+n_3-n_1\equiv N+1}
is an integer, which we denote by $N+1$. 

As described in Appendix B, the residue $R_{3,2,1}$ \resthree\ vanishes for $N<0$. 
Thus, the decay amplitude $1\to 2+3$ vanishes in this case. For $N\ge 0$ one finds
(using  results from the Appendices):
\eqn\roottth{\eqalign{
&R_{3,2,1}={k^2\over 2\pi}\nu(k)^{j_1+j_2+j_3+1}S(2;3) S(3;2)\times\cr
&\prod_{i=1}^N{\Gamma(-{i\over k})\Gamma({2j_1+1-i\over k})\over
\Gamma(1+{i\over k})\Gamma(1-{2j_1+1-i\over k})}
\prod_{i=0}^N{\Gamma(1-{2j_2+1+i\over k})\Gamma(1-{2j_3+1+i\over k})\over
\Gamma({2j_2+1+i\over k})\Gamma({2j_3+1+i\over k})}~,\cr}}
where 
\eqn\sssooo{\eqalign{
&S(2;3)=\sum_{n=\max \{0, N+1-n_2\} }^{\min\{N, n_3-1\}}
{N\choose n}{(-)^{n_3-1-n}\over (n_3-1-n)! (n_2-1+n-N)!}\times\cr
&\prod_{i=0}^{n_3-n-2}(2j_3+n_3+N-n-i)!
\prod_{i=0}^{n_2+n-N-2}(2j_2+n_2+n-i)!~.\cr
}}
$R_{3,2,1}$ \resthree, \roottth\ times the contribution of 
$\IR^{d-1,1}\times LG(W=F)$ to the first line of \threere\ gives
the un-normalized on-shell three point coupling. To get the normalized 
on-shell matrix element $^{\rm out}\langle 3,2|1\rangle^{\rm in}$ one has 
to divide it by $\prod_{i=1}^3{4m_i\over k}R_{n_i,m_i}^{1\over2}$
(see \mompole, \kmuM, \RRR),
\eqn\ththth{^{\rm out}\langle3,2|1\rangle^{\rm in}=
\langle W_{\Delta_3,\bar\Delta_3}e^{ik_3\cdot x}
W_{\Delta_2,\bar\Delta_2}e^{ik_2\cdot x}
G_{-{1\over2}}\bar G_{-{1\over2}} W_{\Delta_1,\bar\Delta_1}e^{ik_1\cdot x}\rangle
{R_{3,2,1}\over \prod_{i=1}^3{2(2j_i+1)\over k}R_{n_i,m_i}^{1\over2}}~.}
To recapitulate, we computed the three point coupling of the on-shell 
states described in section 3 for the case where the momentum around
the cigar is zero, $m_i=\bar m_i$, and the winding number is conserved, 
$\sum m_i=0$. An incoming state corresponding to \inone\ can decay into 
two states \outtwothree\ if and only if
\eqn\boundj{j_1\ge j_2+j_3+1~.}
This selection rule is in fact a consequence of the underlying $SL(2)$
symmetry, \ie\ the fact that for principal discrete series representations,
\eqn\prinjjj{|j_2\rangle\otimes |j_3\rangle=\sum_{j_1\ge j_2+j_3+1}|j_1\rangle~.} 

An interesting question is whether the matrix element \ththth\ has singularities
for different observables. The only possible source of such singularities
is $R_{3,2,1}$ \roottth. The first factor in $R_{3,2,1}$, $\Gamma(-{i\over k})$,
is never singular since $i\le N$ and by using the definition of $N$, \jojtjt,
and the bounds \regone\ on the $j_i$ one can show that $N<(k-1)/2<k$. The second
factor in \roottth\ is likewise regular since $j_1>N$ \regone, \jojtjt, so
$2j_1+1-i>0$ for $i\le N$. The two remaining $\Gamma$ functions in the numerator 
of \roottth\ can similarly be shown to be finite for values of $j_1, j_2, j_3$
which satisfy the unitarity bound \regone. Thus, the decay amplitude \ththth\
is always finite for physical states.

\newsec{Discussion}

The $d$-dimensional Double Scaled Little String Theory defined in \gk\
has the following properties:
\item{(1)} A Hagedorn density of states in a theory without gravity.
\item{(2)} The spectrum of the theory \massgen\ includes
some massless states, Kaluza-Klein type excitations with
masses of order $M_s/\sqrt{k}$, and a stringy spectrum which
starts at masses of order $M_s$.
\item{(3)} The theory has well defined off-shell Green's functions. The analytic
structure of these correlation functions is very reminiscent of local
QFT. In particular, the two point functions have poles corresponding to 
the states \massgen. Moreover, LSZ reduction applies to the three point 
functions and can be used to study on-shell S-matrix elements. 
\item{(4)} The theory is weakly coupled at moderate energies.
The weak coupling approximation breaks down both at high energies
and in some cases at very low ones.  

\noindent
The last point is related to a potentially puzzling aspect of our analysis.
Both here and in \gk\ it was observed that to obtain sensible unitary
amplitudes from the weak coupling analysis of DSLST one needs to restrict
the observables in external legs of different correlation functions. 
Examples include the bound \regone\ which is necessary for unitarity
of the two point functions, and \badpoles\ that is needed to make sense of
three point functions\foot{In fact, as mentioned above, these
bounds are needed in string theory on $AdS_3$ as well, for similar
reasons.}. These bounds cut off the allowed values of off-shell momenta in
the Euclidean regime. For example, the stress-tensor of DSLST
is described in the bulk theory as an on-shell graviton which
is only defined for $k_\mu^2<0$. 

The above bounds on momenta are clearly inconsistent with analyticity
of off-shell Green's functions (it should be emphasized that on-shell
S-matrix elements of the states \massgen\ {\it are} analytic in the
Mandelstam invariants). LST is a non-local theory, and it is
not clear apriori whether it should satisfy off-shell analyticity (see
\aaabbb\ for a recent discussion), but assuming
that it does (as is perhaps suggested by property (3) above), the bounds 
must be reinterpreted. We would like to propose that these bounds are
associated with the breakdown of the weak coupling expansion in DSLST. 
String loops are naively suppressed by powers of $1/\chi$ \scvar, but
it is possible that as one approaches the boundaries of the region
\regone, say, loop corrections become large and invalidate the
perturbative analysis. Thus, at least some of the bounds \regone, \badpoles\ 
might not be bounds on observables in correlation functions but rather
on the reliability of the perturbative analysis of
various correlation functions. It would clearly be interesting to
compute loop corrections to the two and three point functions we considered
here and check this directly. 

For the stress-tensor of DSLST described above, string perturbation theory 
breaks down at long distances ($k_\mu^2=0$). It is
perhaps not surprising that the bulk description breaks down
in this regime, since the ``boundary theory'' (low energy DSLST)
is free at low energies -- \eg\ in the six dimensional example
it is the free theory on $k$ separated $NS5$-branes. In the $AdS/CFT$
correspondence  when the boundary theory is free in the IR,
the bulk theory typically becomes strongly coupled there. Usually, the strong
coupling is associated with large curvatures in the bulk geometry
rather than large corrections from string loops, but we have
to recall that the near-horizon geometry of $k$ $NS5$-branes,
which is a multi-center CHS solution \chs, is related to the
geometry \stvcsix\ by T-duality, which smooths out the large
curvature of the multi-center solution, apparently at the price
of introducing large string loop corrections.

Unlike anti-de-Sitter space, string theory in
asymptotically linear dilaton backgrounds of the sort studied in \abks\
and here has observables corresponding to wave-functions that are 
$\delta$ function normalizable in the radial ($\phi$) direction. 
These observables are needed to construct off-shell operators
in LST with arbitrarily large negative $d$-dimensional momentum
$k_\mu^2$. The role of these
observables in the theory is puzzling, since they seem to correspond
to fluctuating fields in LST and not to external sources, whose norm
should diverge from large $\phi$. Excluding them seems to put a lower
bound on the allowed values of $k_\mu^2$ for each observable, which 
violates off-shell analyticity. Note that:
\item{(1)} The issue of the physical nature of the $\delta$ function normalizable
operators is separate from the discussion of the perturbative unitarity
bounds above. In particular, it should not involve strong coupling effects
in the ``bulk'' string theory. Computing the tree level four point function of 
the observables studied here should shed light on the role of the $\delta$ function
normalizable operators in the theory.
\item{(2)} The fate of the $\delta$ function normalizable observables
is related to another open problem. In \gk\ and here we discussed the
physics of states with a discrete spectrum of masses in $d$ dimensions.
The bulk string theory has in addition a continuum of states, which show
up in the off-shell correlation functions as branch cuts at values of
$k_\mu^2$ corresponding to the $\delta$ function normalizable states.
The fate of this continuum in LST is hence tied to these observables. 

Another set of puzzles is related to the maximally supersymmetric
six dimensional case. Consider \eg\ $k$ $NS5$-branes in IIB string
theory. We studied the system in a regime where the fivebranes
are separated such that the energy of the ground states of D-strings
that stretch between the fivebranes, which is of order $M_W$,
is much larger than the other scales in the problem, $M_s$, $M_s/\sqrt k$.
One might expect that the fivebranes would not interact in this limit,
and the full theory would split into a set of $k$ decoupled theories
living on the different fivebranes. It is believed that there is
no non-trivial theory on a single fivebrane; therefore, one would expect
LST to be trivial for $M_W\gg M_s$. We nevertheless find a non-trivial
theory with a complicated $k$ dependent spectrum of masses \massgen\
and interactions (the simplest of which -- the three point couplings -- 
are described in section 4).

An example of this problem was mentioned at the end of section 3. The analysis
of the two point functions in section 3 leads one to conclude that the
operators ${\rm Tr}\, A^i$ \aabb\ create massless single particle states, 
which therefore carry charge $i=2,3,\cdots, k$ under the unbroken $SO(2)$
symmetry of the vacuum. At low energies one expects to find a $U(1)^{k-1}$
gauge theory with 16 supercharges in which $A$ is a diagonal matrix of
scalar fields, whose eigenvalues create single particle states with
$SO(2)$ charge $1$. Thus, the spectrum of massless particles we find
seems to disagree (as far as the $SO(2)$ quantum number is concerned)
with what one expects from gauge theory.
 
One possible resolution of the above discrepancies is that there is
a subtlety that has been ignored in the relation of string theory on
the manifold \stvcsix\ and the theory on separated fivebranes. For example,
it could be that in the fivebrane theory $A$ has an expectation value as well
as $B$, and the $SO(2)$ invariance is restored by summing in some
way over different expectation values. 

Another possibility is that the identification of the fivebrane system
is correct, and our results point to interesting strong coupling effects 
that exist in this theory at the string scale $M_s$, even when $M_W$ is
very large. This is needed to explain the R-charge spectrum of the massless 
particles and might also help explain the spectrum \massgen. Imagine turning 
on $M_W$ continuously in the theory of $k$ initially coincident fivebranes.
For $M_W\ll M_s/\sqrt k$, the naive gauge theory picture is clearly correct
at low energies, and the low lying spectrum includes $k-1$ 
scalar fields with $SO(2)$ charge $1$. As $M_W$ increases and eventually
passes $M_s$, the theory develops a strong coupling regime at energies
$M_s/\sqrt k<E<M_W$. Thus, it is possible that some operators which
are irrelevant at weak coupling become relevant in this energy regime\foot{Such
operators are known as ``dangerously irrelevant'' in the general theory
of the Renormalization Group.},
and cause the theory to flow to a different fixed point in which the
number of massless degrees of freedom is the same (due to supersymmetry)
but they have different quantum numbers. Examples of such behavior
are known to occur in supersymmetric gauge theories (see \eg\ 
\ref\kss{D. Kutasov, A. Schwimmer and N. Seiberg, hep-th/9510222,
Nucl.Phys. {\bf B459} (1996) 455.}).

\bigskip
\noindent{\bf Acknowledgements:}
We thank O. Aharony, M. Berkooz, S. Elitzur, O. Pelc and A. Zamolodchikov for 
useful discussions. We also thank the ITP in Santa Barbara for hospitality 
during the course of this work. This research was supported in part by NSF 
grant \#PHY94-07194. The work of A.G. is supported in part by the Israel 
Academy of Sciences and Humanities -- Centers of Excellence Program, and by
BSF -- American-Israel Bi-National Science Foundation. D.K. is supported
in part by DOE grant \#DE-FG02-90ER40560.

\appendix{A}{Some useful formulae}

\eqn\gampole{\Gamma(a+1)=a\Gamma(a)~;\;\;\;
\Gamma(-n+\epsilon)={(-)^n\over\epsilon n!}+O(1)~, \qquad n=0,1,2,\cdots}
\eqn\abnm{\int d^2x|x|^{2a} x^n|1-x|^{2b}(1-x)^m=\pi 
{\Gamma(a+n+1)\Gamma(b+m+1)\Gamma(-a-b-1)\over\Gamma(-a)\Gamma(-b)
\Gamma(a+b+m+n+2)}~, \qquad n,m\in Z~.}
\eqn\gggjjj{\eqalign{
G(j)=&G(-j-1-k)~,\cr
G(j-1)=&{\Gamma(1+{j\over k})\over \Gamma(-{j\over k})}G(j)~,\cr
G(j-k)=&k^{-(2j+1)}{\Gamma(1+j)\over \Gamma(-j)} G(j)~.\cr
}}

\appendix{B}{On-shell three point couplings}

For $N\ge 0$ \jojtjt, the factor $G(j_1-j_2-j_3-1)$ in \djk\ has a pole (see 
\recrel, \jojtjt). The integral \jmjmjm\ supplies the other two poles
in \resthree. It can be evaluated by noting that the integrand of
\jmjmjm\ is a polynomial in $|1-x_2|$. One expands
\eqn\expxtwo{|1-x_2|^{2N}=\sum_{n,m=0}^N {N\choose n}{N\choose m}(-)^{n+m}
x_2^n\bar x_2^m~,} 
and uses \abnm\ to evaluate \jmjmjm. Requiring that this integral has poles
at the locations \outtwothree\ constrains $n,m$ in \expxtwo\ to lie
in the interval $N+1-n_2\le n,m\le n_3-1$ (and of course \expxtwo\ 
$0\le n,m\le N$). After some algebra, this leads to \sssooo. 
The other factors in \roottth\ arise from \djk\ by substituting 
\inone, \outtwothree\ and using \recrel.

To show that the on-shell matrix elements vanish for $N<0$, one first notes
that the function $D(j_3, j_2, j_1;k)$ \djk\ contributes no
poles in this case. The factor $G(N)$ in \djk\ is non-singular, since for
negative arguments $G(j)$ does not have poles for $j>-k-1$ and one can show 
that in the regime \regone, $N>-k-{1\over2}$. The factor $G(-j_1-j_2-j_3-2)$
can become singular for $j_1+j_2+j_3\ge k-1$, however, as explained in section
4 (see \badpoles) this regime is excluded by unitarity, or alternatively applicability 
of weakly coupled string theory. The remaining two factors in the numerator
of \djk\ have some poles, but those correspond to processes like $2\leftrightarrow 1+3$
or $3\leftrightarrow 1+2$.  

Thus, to get a non-zero matrix element for $N<0$ one needs to generate all
three poles \resthree\ from the integrals \jmjmjm. An explicit calculation
shows that one can get at most two.

\listrefs
\end